\documentstyle[preprint,aps,psfig]{revtex}
\tightenlines 
\begin{document}
\newcommand{\approxge}{\stackrel{>}{~}} \draft \title{Memory in
2D heap experiments}

\author{Junfei Geng, Emily Longhi, and R.~P. Behringer}
\address{Department of
Physics and Center for Nonlinear and Complex Systems, Duke University,
Durham NC, 27708-0305, USA}

\author{D. W. Howell} \address{Materials Science Division, Argonne
National Laboratory, 9700 S. Cass Ave., Argonne, IL 60439}

\date{\today}

\maketitle

\begin{abstract}

The measurement of force distributions in sandpiles provides a useful
way to test concepts and models of the way forces propagate within
noncohesive granular materials.  Recent theory\cite{bouchaud,wittmer,cates_98} by Bouchaud et
al. implies that the internal structure of a heap (and therefore the
force pathway) is a strong function of the construction history.  In
general, it is difficult to obtain information that could test this idea from 3D granular experiments except at boundaries.  However,
2D systems, such as those used here, can yield information on forces and particle arrangements in the interior of a sample. We obtain position and force information
through the use of photoelastic particles. These experiments show that
the history of the heap formation has a dramatic effect on the
arrangement of particles (texture) and a weaker
but clear effect on the forces within the sample. Specifically, heaps
prepared by pouring from a point source show strong anisotropy in
the contact angle distribution.  Depending on additional details, they
show a stress dip near the center.  Heaps formed from a broad source
show relatively little contact angle anisotropy and no indication of a
stress dip.
\end{abstract}

\pacs{PACS numbers: 46.10.+z, 47.20.-k} 

Granular materials are of great interest for their technical relevance
and for the host of modeling challenges that they
present\cite{jaeger_96}.  Transporting and processing agricultural
grains, coal, and pharmaceutical powders represent only a few applications. Avalanches and mudslides are
important natural phenomena that involve granular materials.  Our
understanding of granular flow, mixing, and even static behavior is
still a open subject.  In this work, we focus on the last of these.

\begin{figure}[htb]
\center{\parbox{3.375in}{\psfig{file=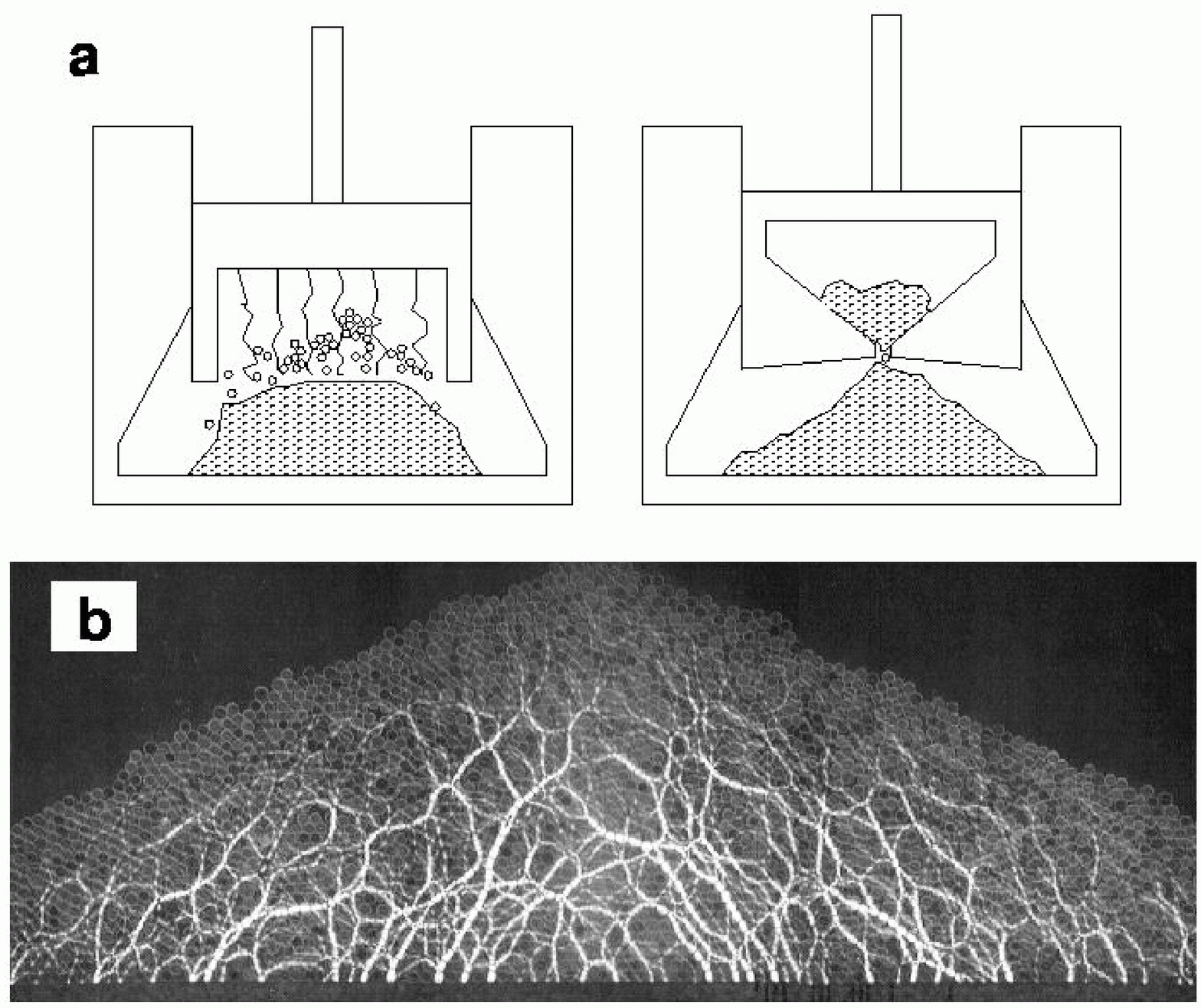,width=3.375in}}}
\caption{{\bf Lower section}: Setup of two-dimensional pile (height $\sim
30$ cm and base length $\sim 130$ cm ) of photoelastic disks created by a
localized-source procedure. The pile is viewed between crossed
polarizers allowing one to see the underlying force structure. Bright regions correspond to the force chains. {\bf Upper
section}: Deposition procedures for triangular piles. Left picture
shows raining technique, right shows localized procedure. See text for
more details.}
\label{fig:heap}
\end{figure}

Static or slowly evolving granular systems are dominated by stress chains, long filamentary structures corresponding
to the paths along which the majority of the force is
carried\cite{chains}.  In the photoelastic image of
Fig.~\ref{fig:heap}b, the bright disks are part of the force chains in
a 2D realization of a sandpile. The formation of these chains plays an
important role in the final static state of a granular system.  One
reason for this history dependence is that the solid-on-solid (SoS)
friction between individual particles is typically indeterminate.
That is, under static conditions, the tangential frictional force at a
contact, $F_T$, can be anywhere in the range $|F_T| \leq
\mu |F_N|$, where $\mu$ is the ordinary SoS friction coefficient and
$F_N$ is the normal contact force.  For collections of stiff frictional grains, it is not always possible to determine
the forces at the contacts solely from the positions of the grains\cite{duran}.

Recently, several authors have proposed/discussed new continuum models
for stresses within sand piles, including the Oriented
Stress Linearity (OSL) model\cite{cates_98},
which is predicated on a picture of how force chains are frozen into a
granular system during its formation, thus creating inherent textures. This model was created in particular to better understand reported stress minima\cite{smid_81,jotaki_79} under the center of
some sand piles. There are other models for granular
statics\cite{nedderman,savage_98,goddard_98}, some of which also can predict a stress dip, including `Incipient Failure Everywhere' (IFE) which assumes that the heap is everywhere at the point of Coulomb failure and elasto-plasticity
models. The stress dip has also been studied via discrete
element models\cite{matuttis,liffman} (DEM) modeling techniques, including Molecular Dynamics (MD) and Contact Dynamics\cite{moreau} (CD). In MD models, a stress dip was found\cite{matuttis,liffman} depending on construction history, or in one case\cite{liffman}, if strong segregation occurred for piles formed from a bidisperse collection of
particles. CD calculations\cite{moreau} predict anisotropy in the contact angles between neighboring particles that reflects the construction of the heap.

The existence of force minima under the base of real sandpiles\cite{smid_81,jotaki_79,brockbank_97,savage_98,vanel_99} has
been equally debated\cite{savage_98}. Considerable care must be taken with experiments because perturbations, such as small deflections of the base surface,
may significantly affect the measured profile.
Recent measurements\cite{vanel_99} on 3D conical piles that had minimal deflections of the base, showed a clear dip when the material was poured on the pile from a localized
source, and no dip when the pile was created by pouring from an
extended source. Additional experiments\cite{vanel_99} on long 3D
heaps (mountain chain shape) showed similar but weaker force sensitivity to how the heap was formed.

Here, we present measurements of contact angle and force distributions for 2D heaps prepared by pouring particles from a) a fixed height point-like source,
b) a slowly moving point source and c) an extended source.  There is a
dramatically different arrangement of contact angles for the point
vs. extended sources, even though the heap angles for the two
techniques are essentially indistinguishable. There is also a clear
difference in the force profiles for the various preparation methods.
For the fixed height localized source, the distribution shows a clear
minimum in the force at the heap center.  For the other two methods,
the force distribution shows a broad maximum at the center, and
possibly other structure on a finer scale.      

The experiments were carried out with disks made from a photoelastic
(birefringent under stress) material\cite{PSM}. The sample was a mixture of two disk sizes, one with diameter $= 0.9$ cm ($\sim 500$ disks) and the other with diameter $= 0.7$cm ($\sim 2500$ disks).  The disks were confined between two
Plexiglas sheets, spaced slightly wider than the thickness (0.6 cm) of
the disks (The flat sides of the disks were parallel to the
Plexiglas).  The surfaces of the parallel sheets were lubricated with a
fine powder and typically oriented no more than 3$^o$ from vertical to minimize friction between the sheets and the disks. The heap size was $\sim 130$ cm at the base, and $\sim$ 30 cm high. We built heaps by the three
different pouring techniques, two local and one extended, as sketched
in Fig.~\ref{fig:heap}a.  For the local source, the disks were held in hollow Plexiglas insert between the two confining Plexiglas
sheets.  The insert had an opening $\sim 7$ grain
diameters wide, small enough to be point-like, but wide enough that flow-stopping arches rarely formed. In the fixed height version of the local source,
the filled insert was placed at a fixed height, 57 cm, above the base.  A stopper was then removed from the insert opening, and the disks flowed out. In
the slowly moving version of the point source, we gradually raised the insert, creating a steady slow flow of disks onto the peak of the emerging pile. The extended source also consisted of an
insert, but the central portion contained strips of
either cardboard or Plexiglas.  When the insert was lifted, it produced
a steady rain of particles.  As the heap formed, some of the particles
then avalanched off, and the final heap profile was not perceptibly different
from those formed by the point-source method.  In all cases, we imaged
the final state of the system with video (640 $\times$ 480 pixels). In
order to obtain high resolution, we obtained two sets of three overlapping images
that covered the central region of the heap, $\sim 110$ grain
diameters wide. One set
of images was obtained with the system placed in a circular
polarimeter\cite{heywood} and one without the polarimeter.  In
obtaining the various sets of images, the particles were left
undisturbed for a given realization. The images with polarimeter yielded force information using a novel application of
photoelasticity described elsewhere\cite{howell_99} and the other
yielded the disk positions.  We analyzed the images without the
polarimeter to obtain the particle centers, and hence, the contact
angles. A given production of a heap showed large variations in
the spatial structure of contacts and stress chain network.  In order
to obtain a reasonable statistical average, we carried out 50
realizations of each heap preparation method.

\begin{figure}[htb]
\center{\parbox{3.0in}{\psfig{file=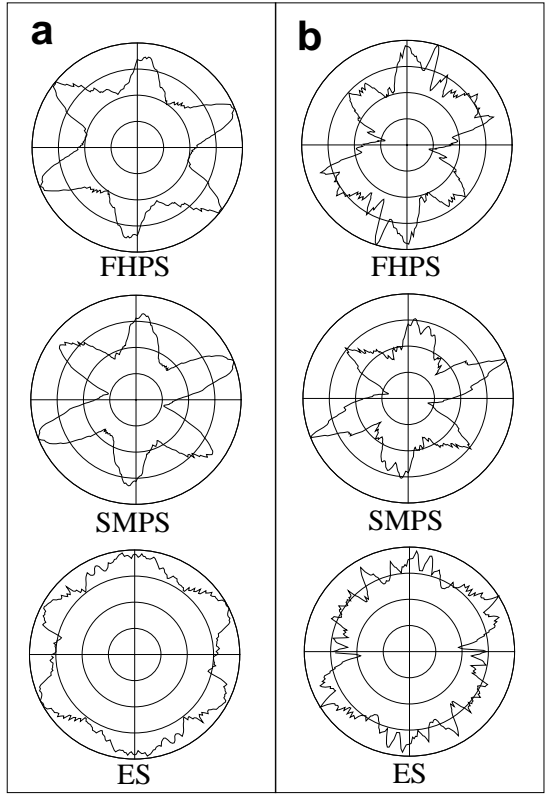,width=3.0in}}}
\caption{{\bf a}: Comparison of mean nearest neighbor angle distributions, $\rho(\theta)$, of all particles for three different pouring techniques: fixed-height point source ({\bf FHPS}), slowly-moving point source {\bf SMPS} and extended source {\bf ES}.  The distributions are only for the left side of the sample typified by Fig. 1b. The $\rho(\theta)$ are given in radial plots and scaled in each plot such that the maximum value is 1 (the outmost circle).  {\bf b}: Comparison of mean nearest neighbor angle distributions for the particles on the force chains for three different pouring techniques. The distributions are for the left side of the sample too.}
\label{fig:dists}
\end{figure}

In order to provide a simple measure of the contact orientation, we evaluated the average angular
distribution of contacts, $\rho(\theta)$, for the left and right sides
(about the vertical center line) separately. The orientation of the particle contacts was dramatically different
for the various preparation methods. We consider a contact to exist if the distance between the centers of two partciles is within $4\%$ of $R_{1} + R_{2}$, where $R_{i}$ are the radii of the two particles. Thus, some of the contacts may not be force-bearing. We only consider data for the left half of the heaps, since similar distributions result for the right side with an
appropriate mirror reflection. In Fig.~\ref{fig:dists}a, we contrast the distributions for the fixed height point source (FHPS), slowly moving point source (SMPS) and the extended sources (ES). For both the localized-source procedures, there is strong anisotropy, and a clear preferred set of orientations.  By contrast, for the extended source procedure, the contacts are much
closer to having an isotropic orientation of contacts.  

We have also analyzed the contacts for only those disks that lie on a stress chain, i.e. the distributions of neighbors of disks that experience a force exceeding the mean. These constitute  $\sim 1/3$ of the disks. We show the corresponding $\rho(\theta)$ for the stress chain disks in Fig.~\ref{fig:dists}b again only for the left half of the heap. In all cases, these large-force-carrying disks have contact angle distributions that break the roughly six-fold symmetry present in the distributions for all disks.  This corresponds to stress chains that are inclined at angles $0 < \theta < \pi/2$, i.e. in such a way as to support the heap.  This compares very well with the CD calculations of Moreau\cite{moreau}.

\begin{figure}[htb]
\center{\parbox{3.0in}{\psfig{file=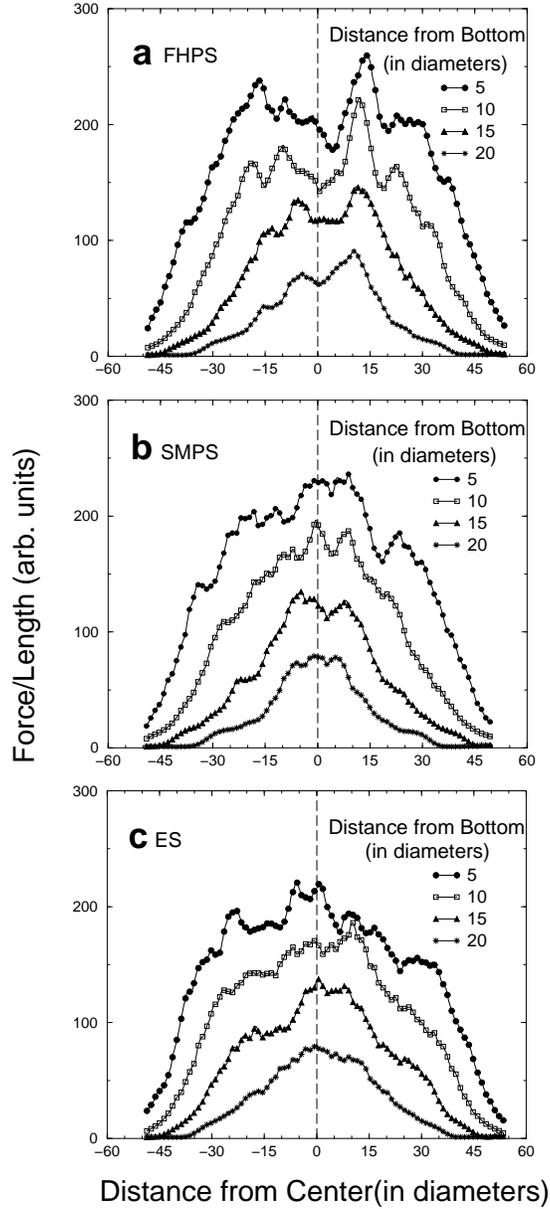,width=3.0in}}}
\caption{Force as a function of horizontal distance along the indicated vertical distances from the bottom of the heap for three different pouring techniques: {\bf a}, fixed-height point source ({\bf FHPS}); {\bf b}, slowly-moving point source ({\bf SMPS}); {\bf c}, extended source ({\bf ES}). The distances are measured in the diameter of the small particle.}
\label{fig:forces}
\end{figure}

The force profiles are affected by preparation history, but not as
 dramatically as the contact angles.  In Fig.~\ref{fig:forces}, we 
show the force (averaged over 50 samples) as a function of horizontal 
distance for various heights from the bottom of the heap, measured in small particle diameters.  The fixed-height local source method clearly yields a force minimum in the center of the heap.  The slowly lifted point source does not show a convincing dip, at least within the scatter of the data.  Similarly, the extended source method shows no dip.

\begin{figure}[htb]
\center{\parbox{3.375in}{\psfig{file=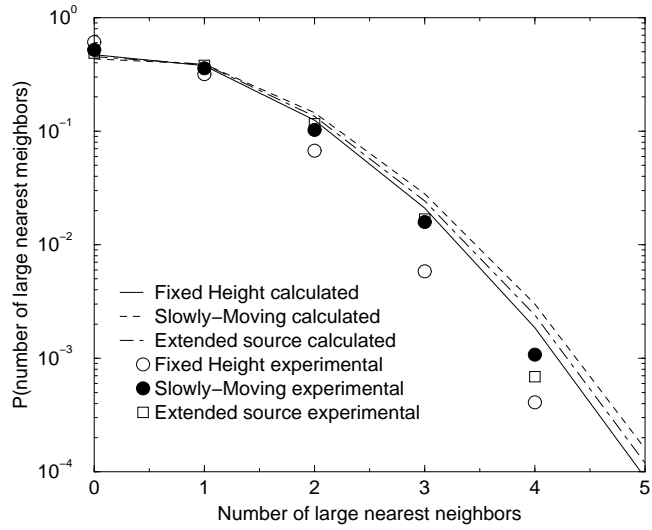,width=3.375in}}}
\caption{Probability distributions for the number of large nearest
neighbors for three different pouring techniques: fixed-height point source ({\bf FHPS}), slowly-moving point source ({\bf SMPS}) and extended source ({\bf ES}). Points are experimental data and the curves are calculated as described in the text.}
\label{fig:nn_dists}
\end{figure}

A final consideration is particle segregation, since significant segregation might lead to a pressure dip\cite{liffman}. In our experiments, we did not
observe the strong segregation assumed in DEM
calculations\cite{liffman}.  This point is documented in
Fig.~\ref{fig:nn_dists}, where we compare experimental distribution
data (points) and calculated random distributions (lines) for the probability $P(n_{l})$ of finding $n_{l}$ large particles next to
any given particle. We estimate $P(n_{l})$ as  $P(n_{l})=\sum_{n=2}^{6} \alpha_{n}P_{n,n_{l}}$ , where $P_{n,n_{l}}$ is the binomial distribution for a total of $n = n_{s} + n_{l}$ contacts ($n_{s}$ the total number of contacts with small particles) and where $\alpha_{n}$ is fraction of particles in the sample with coordination number $n$. We determine the $\alpha_{n}$ separately for each construction technique, hence the different lines in Fig.~\ref{fig:nn_dists}. There is no indication of significant segregation, which may be
because the particles remain in a relatively dense state which limits
the freedom of particles to segregate.

To conclude, these experiments show clearly, and for the first time to
our knowledge, that different methods of heap preparation lead to
dramatically different distributions of contact angles, hence, the
texture, for physical granular systems.  Knowledge of this texture is
then crucial for predicting static granular states.  Although we do
not determine information on the vector forces at frictional contacts, it is
important to emphasize that the most important difference between the
various filling techniques is the amount of disorder in the contact
orientations.  This does not seem to depend on segregation effects.

{\bf Acknowledgments} The work was supported by the
National Science Foundation under Grant DMR-9802602, and DMS-9803305,
and by NASA under Grant NAG3-2372.

\newpage

\end{document}